%% file: main.tex
\documentclass[sigconf]{acmart}

\def\BibTeX{{\rm B\kern-.05em{\sc i\kern-.025em b}\kern-.08emT\kern-.1667em\lower.7ex\hbox{E}\kern-.125emX}}

\usepackage{latexsym}
\usepackage{tikz}
\usetikzlibrary{arrows,shapes,automata, positioning}

\usepackage{ifthen}
\usepackage{amsmath}
\usepackage{multirow}

\usepackage{url}
\usepackage{booktabs}
\usepackage{float}
\usepackage{blindtext}
\usepackage[english]{babel}

\usepackage{amssymb}
\usepackage{color, colortbl}

\usepackage{graphicx}
\usepackage{subcaption}
\usepackage[colorinlistoftodos]{todonotes}
\usepackage[tikz]{bclogo}
\usepackage{fancybox}
\usepackage[belowskip=4pt,aboveskip=4pt]{caption}
\definecolor{Gray}{gray}{0.9}
\definecolor{LightCyan}{rgb}{0.88,1,1}

\setcopyright{acmcopyright}
\copyrightyear{2019}
\acmYear{2019}
\acmConference[CIKM '19]{The 28th ACM International Conference on Information and Knowledge Management}{November 3--7, 2019}{Beijing, China}
\acmBooktitle{The 28th ACM International Conference on Information and Knowledge Management (CIKM '19), November 3--7, 2019, Beijing, China}
\acmPrice{15.00}
\acmDOI{10.1145/3357384.3358145}
\acmISBN{978-1-4503-6976-3/19/11}

\begin{document}
\fancyhead{}


\title{Modeling Long-Range Context for \\ Concurrent Dialogue Acts Recognition}

%
\author{Yue Yu}
\authornote{Both authors contributed equally to this research.}
\email{yy476@georgetown.edu}
\orcid{0000-0003-0566-9304}
\affiliation{%
  \department{Department of Computer Science}
  \institution{Georgetown University}
}

\author{Siyao Peng}
\authornotemark[1]
\email{sp1184@georgetown.edu}
\affiliation{%
  \department{Department of Linguistics}
  \institution{Georgetown University}
 }

\author{Grace Hui Yang}
\email{huiyang@cs.georgetown.edu}
\affiliation{%
  \department{Department of Computer Science}
  \institution{Georgetown University}
}

%


%
%

%



%

\begin{abstract}

In dialogues, an utterance is a chain of consecutive sentences produced by one speaker which ranges from a short sentence to a thousand-word post.
When studying dialogues at the utterance level, it is not uncommon that an utterance would serve multiple functions. For instance, ``Thank you. It works great.'' expresses both gratitude and positive feedback in the same utterance. Multiple dialogue acts (DA) for one utterance breeds complex dependencies across dialogue turns. Therefore, DA recognition challenges a model's predictive power over long utterances and complex DA context. We term this problem Concurrent Dialogue Acts (CDA) recognition. 
Previous work on DA recognition either assumes one DA per utterance or fails to realize the sequential nature of dialogues. 
In this paper, we present an adapted Convolutional Recurrent Neural Network (CRNN) which models the interactions between utterances of long-range context. Our model significantly outperforms existing work on CDA recognition on a tech forum dataset. 

\end{abstract}

\maketitle

\input{intro}
\input{related}

\section{Task Definition}

The task is defined as a CDA recognition problem where for each utterance $u_t$ (the $t$-th utterance) in a dialogue,
we predict a subset of DA labels $y_t$ that describes the functionality of the utterance from a candidate set of DA labels $\mathcal{L} = \{l_1, l_2,...,l_c\}$.
For a dialog with $s$ utterances, the inputs to the algorithm is $\mathcal{U} = \{u_1, u_2,...,u_s\}$, and the output is $\mathcal{Y}=\{y_1, y_2,...,y_s\}$, where $y_t$ is the annotated DA label set for $u_t$, in which $y_t = \{y_t^{1}, y_t^{2},...,y_t^{c}\}$. Here, $y_t^{j} = \{1, 0\}$ denotes whether the $t$-th utterance of the dialog is labeled with DA label $l_j$ or not. When $\sum_{j=1}^c y_t^j > 1$, we say  CDAs are recognized. Given a dialogue $\mathcal{U}$, the goal is to predict the DA sequence $\mathcal{Y}$ from the text.

\begin{figure*}[t]
    \centering
    \includegraphics[trim={7.9cm 5.9cm 7.8cm 6.2cm}, width=0.7\textwidth, clip]{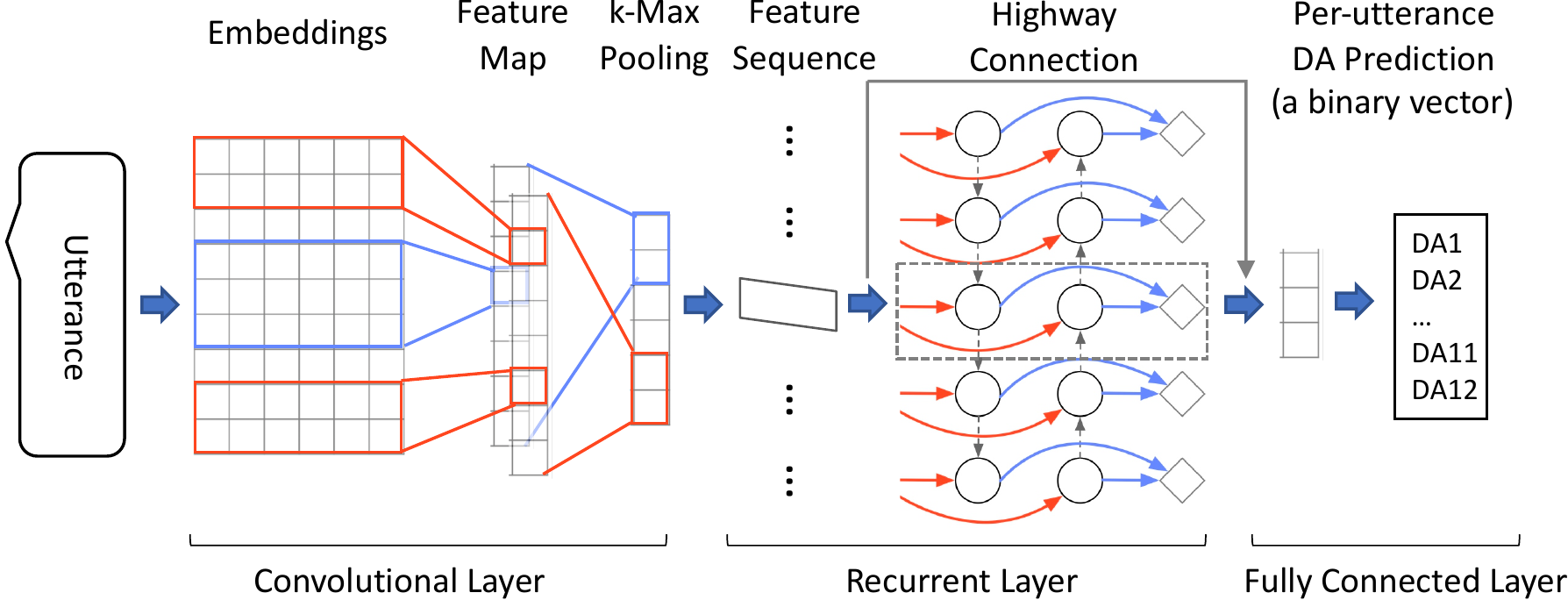}
    \caption{The proposed CRNN model architecture.}
    \label{fig:model_architecture}
    \vspace{-4mm}
\end{figure*}

\section{The Proposed Approach}

The challenge of this task lies in the complexity of dialogue structures in human conversations where an utterance can express multiple DAs. In this work, we improve CDA recognition with an adapted CRNN which models the interactions between long-range context.



\subsection{Convolutional Layer}

The base of our architecture is a CNN module similar to \citet{kim2014convolutional}. The module works by `sliding' through the embedding matrix of an utterance with various filter sizes to capture semantic features in differently ordered n-grams. A convolution operation is denoted as
\begin{align}
k_i &= \tanh(\mathbf{w} \cdot \mathbf{x}_{i:i+d-1}+b_k)
\end{align}
where $k_i$ is the feature generated by the $i$-th filter with weights $\mathbf{w}$ and bias $b_k$. This filter of size $d$ is applied to an embedding matrix, which is the concatenation from the $i$-th to the $(i+d-1)$-th embedding vectors.  This operation is applied to every possible window of words in an utterance of length $n$ and generates a feature map $\mathbf{k}$.
\begin{align}
\mathbf{k} &= [k_1, k_2,...,k_{n-d+1}]
\end{align}

\subsection{Dynamic $k$-Max Pooling}

A max-over-time pooling operation \citep{kim2014convolutional} is usually applied over the feature map and takes the maximum value as the feature corresponding to this particular filter. The idea is to capture the most important features of an utterance. However, this mechanism could be problematic when the utterance is long. We experimented with Dynamic $k$-Max Pooling \citep{liu2017extrememultilabel} to pool the most powerful features from $p$ sub-sequences of an utterance with $m$ words. This pooling scheme naturally deals with variable utterance length.
\begin{align}
p(\mathbf{k}) &= \left[\max \left\{\mathbf{k}_{1 : \lfloor\frac{m}{p}\rfloor}\right\}, \ldots, \max \left\{\mathbf{k}_{\lfloor m-\frac{m}{p}+1\rfloor : m}\right\}\right]
\end{align}


\subsection{Recurrent Layer}
Based on the local textual features extracted from each utterance, a bidirectional RNN is applied to gather features from a wider context for recognizing the DAs in the target utterance, $u_t$. We experimented with two variations of RNN:  LSTM \cite{zaremba2014learning} and GRU \cite{cho2014learning}, both of which utilize gating information to prevent the vanishing gradient problem. GRU is constructed similarly to LSTM but without using a memory cell. It exposes the full hidden state without any control, which may be computationally more efficient. We experimented with both since it is difficult to predict which one performs better on our task.



\subsection{Highway Connection}
Although LSTM and GRU help capture wider context, the network training becomes more difficult with the additional recurrent layers. Inspired by the highway networks \cite{sanders2005highway}, we propose to add a highway connection between the convolutional layer and the last fully connected layer. With this mechanism, the information about the target utterance, $u_t$, can flow across the recurrent layer without attenuation. The last fully connected layer learns from the outputs of both recurrent and convolutional layers.

\section{Dataset}\label{sec:dataset}
We use the MSDialog-Intent dataset \cite{qu2019user} to conduct experiments. In the dataset, each of the 10,020 utterances is annotated with a subset of 12 DAs.
The abundance of information in a single utterance (avg. 72 tokens/utterance) breeds CDA (avg. 1.83 DAs/utterance). We observe a strong correlation between the number of DAs and utterance length, which necessitates a CDA model for forum conversations.

The dataset includes plenty of metadata for each utterance, e.g., answer vote and user affiliation. For generalizability, our model only incorporates textual content of the dialogues. Besides, unlike \citet{qu2019user}, we keep all the DA annotations in the dataset to preserve the meaningful DA structures within and across utterances.\footnote{\label{note1}\citet{qu2019user} simplified the task by removing DA labels from certain utterances and reducing rare DA combinations to top $32$ frequent ones. We re-ran \citet{qu2019user}'s model on the dataset preserving all DA combinations.}

\input{experiment.tex}

\section{Acknowledgements}
This research is supported by U.S. National Science Foundation IIS-145374. Any opinions, findings, conclusions, or recommendations expressed in this paper are of the authors, and do not necessarily reflect those of the sponsor.

\bibliography{citation}
\bibliographystyle{ACM-Reference-Format}

\end{document}

%% file: intro.tex
\section{Introduction}\label{sec:intro}

An utterance is a sequence of sentences that are produced by the same speaker in his or her turn in a dialogue. Dialogue Acts (DA) aim to capture the functionality of these utterances.
 Recognizing DAs can benefit a dialogue system in many ways. 1)  A pre-trained DA model can reduce the number of utterance annotations required to train a dialogue system \citep{kumar2018dialogue2}. 2) For retrieval-based dialogue systems, DA modeling can provide meaningful patterns to reduce the search space and thus expedite response selection. Even though DA recognition is a well-studied problem, most contemporary methods concentrate on conversations of short utterances with usually a single DA per utterance, the case of the Switchboard Dialogue Act (SwDA) corpus \citep{jurafsky1997switchboard}.
However, when utterances are long and complex, one DA label would not be sufficient to describe the pragmatic function of an utterance. As a result, multiple DA labels would be required and the approaches on SwDA would not apply. We term our problem Concurrent Dialogue Acts (CDA) Recognition.

Forum discussion is a type of dialogue that breeds CDAs. 
The sample in Figure  \ref{fig:example_conv380} from MSDialog-Intent corpus \citep{qu2019user} discusses a technical issue with Microsoft Bingo among two users ($U1$ \& $U2$) and two agents ($A1$ \& $A2$). 
For instance, utterance $\mathit{6}$ is simultaneously labeled as GG for ``Good Luck'' and PA for ``using a Win+Shift+Cursor move'' (DA labels are shown in Table \ref{tab:taxonomy_12DAs}). $59.95\%$ of utterances in MSDialog-Intent are indeed multi-labeled.

\input{12DAs_table.tex}

\input{example_fig.tex}

Another important characteristic of these dialogues is that a speaker can refer to any of the preceding utterances. Thus, it is unfeasible to determine a fixed window size that would incorporate all relevant context. In Figure 
\ref{fig:example_conv380}, utterance $\mathit{6}$ provides a potential answer (PA) to utterance $\mathit{1}$; utterance $\mathit{7}$ provides further details (FD) to the information request (IR) in utterance $\mathit{2}$. Without the capability of capturing a wider range of utterances,  the chance of recognizing DAs in these forum conversations would be reduced.

However, current approaches for DA recognition only attempt to capture either of the two characteristics but not both. 
1) Sequence models, that can capture far-away context information, are often applied to DA sequence labeling. \citet{kumar2018dialogue} build a hierarchical RNN using Bi-LSTM as a base unit and CRF as the last layer; \citet{chen2018DACRFASN} use the CRF-Attentive Structured Network for the same hierarchical nature of DAs. Nonetheless, these models only concern single-label DA datasets (e.g., SwDA) and heavily rely on CRF, which is incapable of predicting multiple labels for each utterance.
2) Another line of research casts DA recognition as a multi-label classification problem to accommodate the CDA scenario.
\citet{qu2019user} apply a CNN-based text classifier proposed by \citet{kim2014convolutional} using a fixed window to represent the context. 
Although capable of classifying utterances with CDAs, \citet{qu2019user}'s model only concerns a strictly-local context range and thus cannot include distant information. 

In this paper, we present a novel neural model that is adapted from Convolutional Recurrent Neural Network (CRNN) to both incorporate the interaction between distant utterances and generalize the DA recognition task to accommodate CDA. Our contributions can be summarized as follows: 1) In our adapted Convolutional Recurrent Neural Network (CRNN) we use the recurrent layers to gather long-range contextual information that are extracted from utterances by the convolutional layers. 2) We further optimize the model by incorporating highway connections and dynamic $k$-max pooling. 3) Our model significantly outperforms the state-of-the-art model by 4.68\% in accuracy and 3.08\% in F-score.

%% file: 12DAs_table.tex
\begin{table}[t]
\small
\caption{Taxonomy of 12 dialogue acts in a tech forum.}
\makebox[0.43 \textwidth][c]{       
\resizebox{0.43 \textwidth}{!}{   
    \centering
    \begin{tabular}{|l|l|l|l|l|l|}

\toprule
\textit{DA} & \textit{Taxonomy} & \textit{\%} & \textit{DA} & \textit{Taxonomy} & \textit{\%} \\

\midrule
  
  GG &  Greetings/Gratitude  & 40.0 & 
 FQ &  Follow-up  Question & 8.6 
  \\
PA &  Potential Answer & 39.7 & 
 NF  & Negative Feedback & 7.6  
\\
FD &  Further Details & 24.8 & 
CQ &  Clarifying Question & 7.5  
\\
 OQ & Original Question & 23.3 & 
RQ &  Repeat Question & 6.1 
 \\
 PF &  Positive Feedback & 10.7 &  
 JK &  Junk & 2.6 
 \\
IR &  Information Request &  10.7 & 
 O  & Others  & 1.5 \\
\bottomrule
 \end{tabular}
 }
 }
\label{tab:taxonomy_12DAs}
\end{table}

%% file: example_fig.tex



\begin{figure}[t]
\vspace{-4mm}

    \centering
$U1$
\setlength{\fboxsep}{3pt}%
\cornersize{.3}
\ovalbox{%
\small
\begin{minipage}{7cm}
\textcolor{blue}{$\mathit{1}$} 
For around a month my settings would disappear ... If somebody know how to fix this , please tell me .  \textcolor{blue}{[ OQ ]}
\end{minipage}}

\vspace{1pt}

\setlength{\fboxsep}{3pt}%
\cornersize{.3}
\ovalbox{%
\small
\begin{minipage}{7cm}
\textcolor{blue}{$\mathit{2}$} 
Hi ... to isolate the issue ... we would like to know the following ... look forward to your response .
    \textcolor{blue}{[ GG IR PA ]}
\end{minipage}}
$A1$

\vspace{1pt}

$U2$
\setlength{\fboxsep}{3pt}%
\cornersize{.3}
\ovalbox{%
\small
\begin{minipage}{7cm}
\textcolor{blue}{$\mathit{3}$} 
I want to reinstall Microsoft bingo for free I have Windows 10 I can't even open my store app ...  \textcolor{blue}{[ CQ FD ]} 
\end{minipage}}

\vspace{-2pt}

$U2$
\setlength{\fboxsep}{3pt}%
\cornersize{.3}
\ovalbox{%
\small
\begin{minipage}{7cm}
\textcolor{blue}{$\mathit{4}$} 
I did n't have this issue before
    \textcolor{blue}{[ FD ]}
\end{minipage}}

\vspace{-2pt}

$U2$
\setlength{\fboxsep}{3pt}%
\cornersize{.3}
\ovalbox{%
\small
\begin{minipage}{7cm}
\textcolor{blue}{$\mathit{5}$} 
There were some issues made
    \textcolor{blue}{[ FD ]}
\end{minipage}}

\vspace{1pt}

\setlength{\fboxsep}{3pt}%
\cornersize{.3}
\ovalbox{%
\small
\begin{minipage}{7cm}
 \textcolor{blue}{$\mathit{6}$} 
... you would also have the ( simpler ) alternative of using a Win+Shift+Cursor move . Good luck ... \textcolor{blue}{[ GG PA ]} 
\end{minipage}}
$A2$

\vspace{1pt}

$U1$
\setlength{\fboxsep}{3pt}%
\cornersize{.3}
\ovalbox{%
\small
\begin{minipage}{7cm}
\textcolor{blue}{$\mathit{7}$} 
...The only changes I have made prior to the issue was the installation of new RAM cards ... Windows updates ...
    \textcolor{blue}{[ FD ]} 
\end{minipage}}

\vspace{1pt}

$U1$
\setlength{\fboxsep}{3pt}%
\cornersize{.3}
\ovalbox{%
\small
\begin{minipage}{7cm}
\textcolor{blue}{$\mathit{8}$} 
Thank you ... This works 
    \textcolor{blue}{[ GG PF ]} 
\end{minipage}}

\vspace{-2pt}

$U1$
\setlength{\fboxsep}{3pt}%
\cornersize{.3}
\ovalbox{%
\small
\begin{minipage}{7cm}
\textcolor{blue}{$\mathit{9}$} 
I manage to fix the situation ... thank you ...
    \textcolor{blue}{[ GG PF ]}
\end{minipage}}

    \caption{Dialogue between two users and two agents.}
    \label{fig:example_conv380}
\end{figure}

%% file: related.tex
\section{Related Work}\label{sec:relate}

CRNN has been widely applied to
classification tasks. In Natural Language Processing (NLP) research including sentence segmentation \citep{treviso-etal-2017-sentenceCRNN}, sentiment analysis \citep{wang-etal-2016-combinationsentiment} and discourse modeling \citep{kalchbrenner2013recurrent}, CRNN achieves the state-of-the-art in single-label classification where for each input, a single output would be predicted. 

Deep learning research outside of NLP proves CRNN suitable for classifications where multiple labels may be assigned to each target. 
\citet{cakir2017polyphonicsound} apply CRNN to Polyphonic Sound Event Detection where multiple sound events can be detected in a timestamp. Similar to single-label CRNNs, features are extracted  by the convolutional layers and further integrated into the recurrent layers to provide context information. The sigmoid activation function is used to make multi-binary decisions for each event activity. 
\citet{choietal2017CRNNmusic} treat music tagging as a multi-label classification task where a subset of genres,
moods, etc. are assigned to each music. In \citet{choietal2017CRNNmusic}'s model, The CNN is responsible for extracting local features and the RNN
presents a temporal summarization. 
These studies substantiate CRNN's applicability to multi-label classification tasks, especially its ability to extract features from the current step and to integrate features from arbitrarily long distance.

In previous dialogue research, \citet{kalchbrenner2013recurrent} and \citet{ kim-etal-2016-exploringdialog} incorporate CRNN into a single-label DA recognition and dialogue topic tracking system. 
Based on this line of research, we argue that CRNN can be applied for CDA recognition with two adaptations: binary cross-entropy loss (BCE) and sigmoid activation. Further experiments also show that adding highway connections and dynamic $k$-max pooling further boosts the performance.

%% file: experiment.tex
\section{Experiments}
\label{sec:experiment}
 In this section, three versions of our proposed model with incremental improvements are evaluated against a CNN baseline \citep{kim2014convolutional} and the state-of-the-art approach for CDA recognition \cite{qu2019user}.
 
 \begin{itemize}
\item \textbf{CNN-Kim\cite{kim2014convolutional}:} One of the first attempts to apply CNN to text classification. The CNN model consists of three convolutional layers with the same filter size.
 \item \textbf{CNN-CR\cite{qu2019user}:}
 The state-of-the-art approach for CDA recognition on the MSDialog-Intent dataset \cite{qu2019user}. The CNN model incorporates context information with a window size of 3.
 \item \textbf{CRNN ($v_1$):} Our base model that  adapts CRNN for CDA recognition using BCE loss and sigmoid activation function.
 \item \textbf{CRNN ($v_2$):} CRNN ($v_1$) with highway connections added between the convolutional layer and the fully connected layer.
 \item \textbf{CRNN ($v_3$):} CRNN ($v_1$) with highway connections and dynamic $k$-max pooling implemented.
 \end{itemize}

 \subsection{Experimental Setup}
 The dataset is partitioned into training, validation, and test sets in the ratio of  8:1:1. Hyper-parameters are tuned with the validation set. For word embedding, we use the publicly available GloVe vectors that were trained on Wikipedia + Gigaword 5 \cite{pennington2014glove}. We find that setting (CNN filters, CNN dropout, RNN units, RNN layers, RNN dropout and $k$) to (100, 0.4, 900, 2, 0.15 and 2) is the best for our model. The BCE is used as loss function and Adam \cite{kingma2014adam} is used for optimization with an initial learning rate, 0.001. The models were trained on four GeForce RTX 2080 Ti GPUs.

\subsection{Metrics}\label{subsubsec:metrics}

Following previous work \cite{qu2019user} on multi-label classification, we adopt label-based accuracy (i.e., Hamming score) and micro-$F_1$ score as our main evaluation metrics. Micro-precision and micro-recall are also reported to assist the analysis. Among all,  accuracy is the only metrics that is on a per utterance basis. Therefore, Student's paired t-test is performed only on accuracy. Other metrics (P, R, $F_1$) provide an overall performance evaluation for all utterances.

\subsection{Results}\label{sec:result}
All proposed adaptations, i.e., highway connections and dynamic $k$-max pooling, contribute to the model, as in Table~\ref{tab:result}.
The CRNN models significantly\footnote{\label{sig}* denotes a statistically significant difference over the best baseline (CNN-CR) with p $\leq$ 0.05, ** for p $\leq$ 0.01, *** for p $\leq$ 0.001 and **** for p $\leq$ 0.0001.} outperform both baselines. 
The model with full adaptation ($v_3$), performs the best across all experiments: achieving the highest accuracy, recall and $F_1$  with LSTM and the highest precision with GRU. The divergence between LSTM and GRU may indicate that the \textit{forget} and \textit{output} gates in LSTM are helpful for the retrieval of related memory but increase the variance of the model.

\input{result_maintable.tex}

\input{casestudy_table.tex}

\begin{figure}[t]
    \centering
    \includegraphics[width=0.45\textwidth]{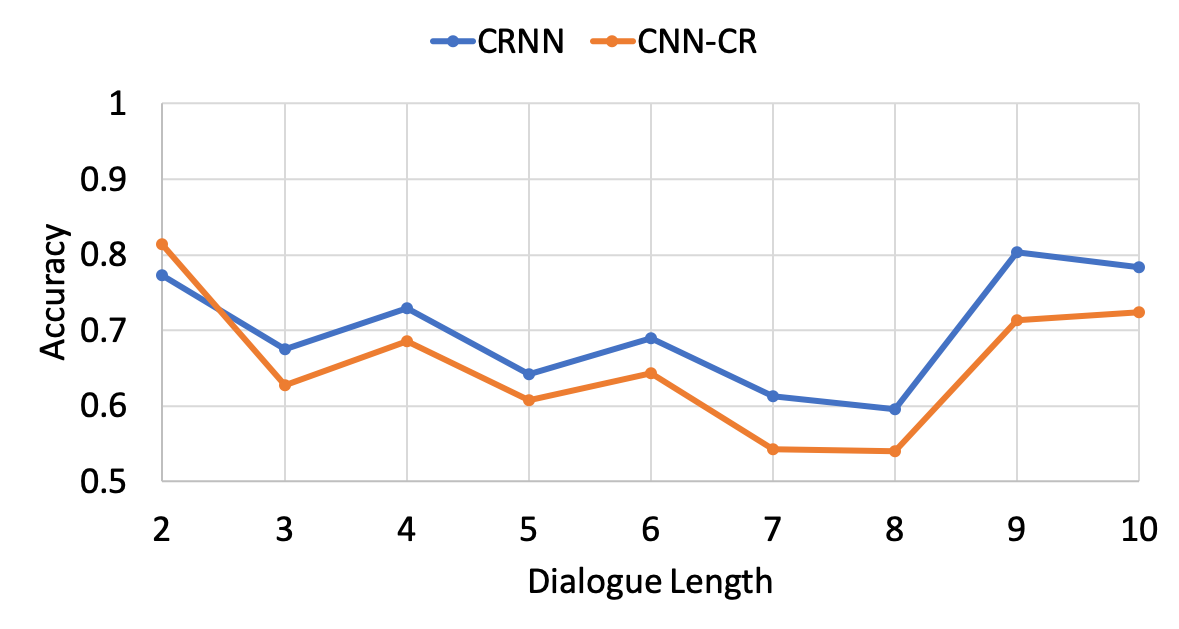}
    \caption{Mean accuracy of CRNN ($v_3$) vs. CNN-CR.}
    \label{fig:compare1}
\end{figure}

Table \ref{tab: casestudy} further exemplifies the capability of our best model in recognizing CDAs. Our model systematically outperforms the state-of-the-art approach, CNN-CR. More specifically, 1) CRNN achieves better mean accuracy for all reference sizes and significantly for sizes $1-3$, which consist of 96.4\% utterances. 2) The average number of DAs predicted by CRNN is the closest to the reference.

Another advantage of CRNN is the ability to handle longer dialogues. Figure \ref{fig:compare1} visualizes the mean accuracy of utterances grouped by dialogue length. Along with the increase of dialogue length, we observe an increasing advantage of CRNN (in blue) over CNN-CR (in orange), especially for dialogues longer than 6 utterances. 

\subsection{A Case study}

Table~\ref{tab: case} shows an example where CRNN and CNN-CR predict differently for utterance $\mathit{5}$.
Utterance $\mathit{5}$ should respond to utterance $\mathit{2}$, establishing a long-range dependency that cannot be captured by CNN-CR's fixed context window. Our CRNN model succeeds in recognizing the distant context for further details (FD). Furthermore, CNN-CR wrongly recognizes utterance $\mathit{5}$ as original question (OQ) since the OQ in utterance $\mathit{1}$  is out of range.

\input{case}

\section{Conclusion}

Our proposed CRNN models for CDA recognition impose fewer restrictions on the structure of DAs and capture textual features from a wider context. The experiment results show that all of the proposed adaptations, i.e., highway connections and dynamic $k$-max pooling, contribute to the model optimization. Our final model significantly outperforms the state-of-the-art approach on a tech forum dataset where the dialogues are packed with complex DA structures and information-rich utterances. Future work will consider a more fine-grained set of DAs for a deeper analysis of CDA modeling. It would also be interesting to annotate antecedent relations among utterances to structure threading dialogue flows.

%% file: result_maintable.tex
\begin{table}[t]
\caption{Performance of CNN-Kim, CNN-CR, and CRNN.\textsuperscript{\ref{sig}} 
}
\centering
\resizebox{0.45 \textwidth}{!}{ 
\begin{tabular}{|l|llll|}
\hline
Models & \multicolumn{1}{l|}{\textbf{Accuracy}} & \multicolumn{1}{l|}{Precision} & \multicolumn{1}{l|}{Recall} & \multicolumn{1}{l|}{$F_1$ score} \\ \hline
 
 CNN-Kim\cite{kim2014convolutional} & 0.5785 & 0.6371 & 0.6745 & 0.6553 
 \\ \hline

CNN-CR\cite{qu2019user}\textsuperscript{\ref{note1}} & 0.6354 & 0.7108 & 0.6952 & 0.7029 

 \\ \hline
CRNN ($v_1$) w/ LSTM & 0.6668* & 0.7238 & 0.7297 & 0.7267 
 \\ \hline
CRNN ($v_1$) w/ GRU & 0.6543* & 0.7056 & 0.7065 & 0.7061  

\\ \hline
CRNN ($v_2$) w/ LSTM & 0.6731**** & 0.7315 & 0.7315 & 0.7315 
\\ \hline
CRNN ($v_2$) w/ GRU & 0.6734** & 0.7280 & 0.7334 & 0.7307 
 \\ \hline
CRNN ($v_3$) w/ LSTM & \textbf{0.6822}**** & 0.7254 & \textbf{0.7422} & \textbf{0.7337}
\\ \hline
CRNN ($v_3$) w/ GRU & 0.6733*** & \textbf{0.7358} & 0.7215 & 0.7286
 \\ \hline

\end{tabular}
}
\label{tab:result}
\vspace{-2mm}
\end{table}

%% file: casestudy_table.tex
\begin{table}[t]
\caption{Mean accuracy and the average number of predicted DAs grouped by the number of reference DAs.\textsuperscript{\ref{sig}} The percentage indicates the frequency of each DA group in the test set.}
\resizebox{0.47 \textwidth}{!}{   
\begin{tabular}{|c|l|l|l|l|l|}
\hline
\multirow{2}{*}{\begin{tabular}[c]{@{}l@{}}\# of \\ ref DAs\end{tabular}} & \multirow{2}{*}{\%} & \multicolumn{2}{c|}{Mean accuracy} & \multicolumn{2}{c|}{Avg. num. of pred DAs} \\ \cline{3-6} 
 &  & CRNN ($v_3$) & CNN-CR & CRNN ($v_3$) & CNN-CR\\ \hline
1 & 36.9   & \textbf{0.7704}** & 0.7126 & 1.44 & 1.44 \\ \hline
2 & 42.8  & \textbf{0.6641}*** & 0.6232 & \textbf{2.02}** & 1.89 \\ \hline
3 & 16.7  & \textbf{0.5596}* & 0.5177 & \textbf{2.56}***  & 2.37 \\ \hline
$\geq$4 & 3.6  & \textbf{0.5618} & 0.5339 & \textbf{2.68} & 2.74  \\ \hline
\end{tabular}
}
\label{tab: casestudy}
\vspace{-2mm}
\end{table}

%% file: case.tex
\begin{table}[t]
\setlength{\baselineskip}{1.5\baselineskip}
\LARGE
\caption{Predictions of CRNN ($v_3$) and CNN-CR on a dialogue with long-range dependencies.}
\resizebox{0.47 \textwidth}{!}{   
\begin{tabular}{|l|l|l|l|l|}
\hline
 \addlinespace[1mm]
 \# & Utterance &  \parbox{0.6cm}{Ref} & {\large \parbox{1.2cm}{CRNN}} & {\large \parbox{0.7cm}{CNN \\ -CR}}  \\  \addlinespace[1mm]
 \hline
 \addlinespace[1mm]
$\mathit{1}$ &  \parbox{8.5cm}{ How can I download Skype for Windows 8.1 ...}
 & {\large OQ}  & {\large OQ} & {\large OQ}   \\  \addlinespace[1mm]
 \hline
  \addlinespace[1mm]
$\mathit{2}$ & \parbox{8.5cm}{ 
Hi ... if you are using a phone running Windows 8.1 ... no longer ... but if ... computer ... you can ...
}  & {\large \parbox{0.6cm}{GG \\ PA}}  &   {\large  \parbox{1.2cm}{GG PA}} & 	{\large \parbox{0.7cm}{GG \\ PA}} \\ \addlinespace[1mm]
 \hline
  \addlinespace[1mm]
\multicolumn{5}{|c|}{\textit{... Utterances $\mathit{3}$ \& $\mathit{4}$ ...}} \\
\hline
  \addlinespace[1mm]
$\mathit{5}$ & \parbox{8.5cm}{ Hi , I am using \textbf{surface tablet} 8.1 Windows ...  tried many times ... this app can't run on this PC} & {\large  \parbox{0.6cm}{FD  \\ FQ \\ GG}} & {\large \parbox{1.2cm}{FD GG \\ PA RQ}} & 	{\large \parbox{0.7cm}{OQ \\ GG}} \\  \addlinespace[1mm]
 \hline

\end{tabular}}
\label{tab: case}
\vspace{2mm}
\end{table}


